\documentclass[conference]{IEEEtran}
\IEEEoverridecommandlockouts
\usepackage{cite}
\usepackage{amsmath,amssymb,amsfonts}
\usepackage{algorithm}
\usepackage{algorithmic}
\usepackage{graphicx}
\usepackage{textcomp}
\usepackage{xcolor}
\usepackage{placeins}
\usepackage{float}
\usepackage{dblfloatfix}
\usepackage[inline]{enumitem}

\def\BibTeX{{\rm B\kern-.05em{\sc i\kern-.025em b}\kern-.08em
    T\kern-.1667em\lower.7ex\hbox{E}\kern-.125emX}}
\begin{document}

\title{Enabling Blockchain Interoperability Through Network Discovery Services}

\author{
\IEEEauthorblockN{Khalid Hassan, Amirreza Sokhankhosh, Sara Rouhani,}
\IEEEauthorblockA{Department of Computer Science, University of Manitoba, Winnipeg, Canada\\
Email: \{hassank2, sokhanka\}@myumanitoba.ca,  sara.rouhani@umanitoba.ca}
}

\maketitle

\begin{abstract}
Web3 technologies have experienced unprecedented growth in the last decade, achieving widespread adoption. As various blockchain networks continue to evolve, we are on the cusp of a paradigm shift in which they could provide services traditionally offered by the Internet—but in a decentralized manner—marking the emergence of the \textit{Internet of Blockchains}. While significant progress has been achieved in enabling interoperability between blockchain networks, existing solutions often assume that networks are already mutually aware. This reveals a critical gap: the initial discovery of blockchain networks remains largely unaddressed. This paper proposes a  decentralized architecture for blockchain network discovery that operates independently of any centralized authority. We also introduce a mechanism for discovering assets and services within a blockchain from external networks. Given the decentralized nature of the proposed discovery architecture, we design an incentive mechanism to encourage nodes to actively participate in maintaining the discovery network. The proposed architecture implemented and evaluated, using the Substrate framework, demonstrates its resilience and scalability, effectively handling up to 130,000 concurrent requests under the tested network configurations, with a median response time of 5.5 milliseconds, demonstrating the ability to scale its processing capacity further by increasing its network size.

\end{abstract}


\section{Introduction}\label{introduction}
The ability to implement a decentralized, secure, yet trustless system of geographically distributed nodes has made blockchain a highly promising technology for various industries. Despite initially being employed for powering Bitcoin \cite{nakamoto2008bitcoin} as the first cryptocurrency, it has seen rising adoption in other sectors such as finance, healthcare, agriculture, and supply chain management \cite{blockchain_banks_dashkevich, blockchain_healthcare_jabbar, xiong_agri, blockchain_supplychain_dutta}. The qualities that make blockchain appealing in these areas include network-specific protocols that define rules for data exchange, governance models that dictate decision-making processes, consensus mechanisms that ensure agreement across nodes, and transaction finality models that guarantee the permanence of validated transactions. However, as blockchain adoption grows,  the need for interoperability between these networks becomes a looming necessity. Despite their benefits, these characteristics are implemented uniquely across networks, which directly contributes to the challenges of blockchain interoperability and often leads to the creation of fragmented data silos \cite{abebe_ibm_iop}.


Currently, most interoperability solutions focus primarily on enabling cross-chain transactions. However, as the blockchain ecosystem evolves, interoperability requirements are expanding beyond transactional capabilities to include arbitrary data sharing—mirroring the functionality of the traditional Internet and advancing the vision of the Internet of Blockchains (IoB) \cite{WeiCrosschain}. While several frameworks, such as Polkadot \cite{wood2016polkadot} and Cosmos \cite{cosmos_whitepaper}, have made notable strides in facilitating cross-chain interactions, they overlook a critical prerequisite: initial network discovery. In practice, discovery mechanisms in existing systems rely on pre-defined lists of known networks \cite{tamIoB}, which are either centrally curated by governing entities—introducing risks of centralization and governance bias—or statically configured and manually maintained by clients \cite{khorasaniAutomatedGateways, wangIntertrust, weaver}.


To address the discovery challenge, this paper proposes an architecture inspired by traditional Web2 DNS to enable inter-blockchain network and service discovery. As a foundational component for realizing IoB, this solution emphasizes dynamic network discovery while maximizing decentralization to reduce governance bias and enhance reliability. The architecture is designed to scale effectively, capable of handling large volumes of records and efficiently resolving queries. This is achieved through a multi-chain design combined with robots incentive mechanisms that leverage off-chain workers, ensuring that beneficiaries actively participate in infrastructure maintenance rather than passively benefiting from it. The proposed system is evaluated using key performance metrics, including query throughput, domain name resolution latency, and the scalability of storage requirements inherent to its multi-chain design.


This study aims to establish foundational groundwork for a more inter-connected blockchain ecosystem, advancing toward the realization of IoB. Accordingly, this paper's contributions are three-fold:
\begin{enumerate}
    \item To the best of our knowledge, this work presents the first architectural and practical realization of a decentralized network discovery and resolution infrastructure for blockchain ecosystems, systematically addressing a previously overlooked yet foundational challenge in achieving scalable and decentralized blockchain interoperability.
    \item We introduce a novel discovery protocol that enables both blockchain network discovery and cross-chain service resolution, thereby significantly enhancing the accessibility and interoperability of decentralized systems.
    \item We design and implement a robust incentive mechanism that promotes active participation in maintaining the infrastructure, discouraging free-riding behavior among beneficiaries.
    \item We conduct a comprehensive evaluation of the proposed architecture under realistic load conditions, demonstrating its scalability, low-latency resolution performance, and resilience to high query volumes, thereby validating its suitability as foundational infrastructure for inter-chain communication.
\end{enumerate}


The rest of this paper proceeds to establish background for the discovery problem and highlight the motivation behind our work in Section \ref{background}. The proposed design is then discussed in Section \ref{design}, followed by detailing the incentive strategy employed in Section \ref{incentive}, leading to the implementation details in Section \ref{implementation} which is followed by design discussions and the implementation's evaluation in Sections \ref{discussions} and \ref{evaluation}. Finally, the paper concludes with a discussion of the achieved results and suggestions for future research directions in Section \ref{conclusion}.

\section{Background and Motivation}\label{background}

\subsection{Interoperability}

The diverse pool of blockchain platforms utilizing different consensus protocols frames interoperability as a layered challenge. Inspired by the OSI model \cite{Zimmermann1980}, the Overledger project \cite{overledger} defines a four-layer model: transaction (ledger storage), messaging (data exchange), filtering and ordering (message validation), and application layers. Similarly, another model \cite{towardsLohachab2021} proposes a seven-layer architecture: de-application, virtual-chain, access, consensus, distributed ledger, message cache, and gateway layers. Despite varying in layer count and terminology, these approaches encompass the same core areas with differences in layer granularity.

This layered approach enables the interoperability problem to be tackled using a divide-and-conquer strategy, enabling tailored solutions for each layer. However, no standard model has been widely adopted, as most solutions are multi-layered and use-case specific \cite{zamyatin_sok_2019}. For instance, the Interledger Protocol (ILP) \cite{interledger} operates across the application, messaging, and networking layers. While it initially depended on conditional channels with Hashed Time-Lock Contracts \cite{lightningNet}, ILP has since shifted to a stateless, packet-switched approach for greater scalability and flexibility.

Since ILP works with different ledgers, it is considered a blockchain-agnostic protocol \cite{wangSurv2023}. Blockchain interoperability solutions can be organized into four categories: Notary schemes, which use trusted entities to manage cross-chain transactions; relays, where one blockchain tracks the state of another (e.g., Polkadot's Parachains that connect to a central Relay Chain for shared security and consensus) \cite{wood2016polkadot}; Hashed Time-Lock Contracts, which require the recipient to acknowledge transactions within a specified timeframe or return funds to the sender\cite{hashlock, timelock}; and blockchain of blockchains, which connect multiple blockchains through shared protocols, such as Cosmos' IBC \cite{cosmos_ibc} and Polkadot's XCMP \cite{polkadotXCMP}, with smart contracts playing a major role in defining the behavioral logic of these solutions \cite{belchior_iop_survey, kannengiesser2020, khanContracts2021}. 

\subsection{Discovery Services}

The current interoperability solutions make strides in enabling cross-blockchain interaction; however they often overlook the crucial problem of blockchain network discovery. These solutions usually rely on a centralized authority to track accessible blockchain networks. While achieving fully decentralized interoperability is a significant challenge \cite{lafourcade2020}, key aspects such as discovery services can still be decentralized, thereby reducing the reliance on centralization. A key objective of blockchain interoperability is to create an Internet of Blockchains (IoB), potentially evolving to an alternative decentralized internet infrastructure \cite{tamIoB, zarrin_blockchain_2021}. This mainly depends on the existence of decentralized discovery services, akin to Web2's DNS \cite{mockapetris_domain_1987}.

The IEEE Standard 3205 \cite{ieee_standard2023} defines protocols for cross-chain data authentication and communication. It does so by organizing interoperability into several layers: application, cross-chain transaction, transport, secure authentication, and data link layers. It proposes a standardized Unified Cross-chain Packet (UCP) structure alongside protocols tailored to each layer. Notably, this standard specifically discusses blockchain network discovery, emphasizing the need for a Blockchain DNS (BCDNS). The BCDNS seeks to provide blockchain networks with globally unique, verifiable identities using a combination of certificate authorities and trust anchors, drawing clear parallels to the traditional DNS system.

To the best of our knowledge, no existing solutions provide discovery services specifically for blockchain networks. However, blockchain technology has been utilized for decentralize DNS. Namecoin \cite{namecoin}, a Bitcoin fork, stores ".bit" domains on the blockchain, decentralizing DNS but facing scalability limitations, such as a 64-character cap on domain names and issues like name-squatting \cite{pataskisUnravelling}. Blockstack \cite{blockstack} addresses scalability by extending Namecoin to the Bitcoin network, utilizing a virtual chain for naming logic that operates independently of blockchain consensus. EmerDNS \cite{emerdns} mitigates name-squatting through the use of variable expiration dates and renewal fees to discourage abusive practices.

Because both Namecoin and Blockstack are built on Bitcoin, they rely on Proof of Work (PoW) for consensus, which imposes significant performance limitations. B-DNS \cite{bdns} addresses these limitations by employing a less resource-intensive Proof of Stake (PoS) consensus mechanism and incorporating a domain index tree to enhance query efficiency. Similarly, DNS-BC \cite{dnsbc} leverages a consortium blockchain to implement a DNS caching system with high accuracy, robust security, and real-time performance. DNS-BC also reduces latency through the DNS over KCP (DoK) protocol and enhances security by utilizing a credibility score that influences the probability of a participating node being selected for a query. 


\subsection{Motivation}

The discovery mechanisms utilized in contemporary blockchain interoperability solutions are centralized, introducing certain limitations and risks. Centralization inherently creates the potential for governance bias, where the controlling entity has full authority over all functionality. This control enables the governing entity to exclude specific networks, preventing them from becoming discoverable, either due to conflicts of interest or strategic considerations. Furthermore, if the entity loses interest or ceases to maintain the service, all blockchain networks reliant on it for discovery are adversely affected -- introducing a single point of failure.

The limitations of current centralized discovery mechanisms extend beyond governance bias and unreliability. In existing interoperability solutions, whether they rely on a centralized repository of known networks or maintain their own local lists, the process of registering new networks typically involves substantial manual effort. This includes governing entities evaluating and approving networks that request access to their service and subsequently updating their repositories. For solutions relying on local lists, administrators must manually modify these lists, a process that often mandates a network restart. These approaches hinder scalability and adaptability, undermining the efficiency of blockchain interoperability.

The aforementioned challenges emphasize the need for a solution that places blockchain discovery at the forefront rather than treating it as an afterthought in interoperability design. The solution presented in this study primarily focuses on the secure authentication layer of interoperability according to the IEEE Standard 3205, addressing the discovery challenges inherent in current approaches. It prioritizes decentralization, ensuring that no single entity controls the network's behavior. The proposed system also introduces dynamic network discovery, moving away from the static and manual mechanisms currently in use. Additionally, the solution is designed with scalability as a core principle, enabling the registration of a large number of networks while maintaining high performance for discovery queries. Using these design principles, this study aims to provide the necessary foundational infrastructure required for inter-blockchain discovery to realize the vision of an Internet of Blockchains.

\section{Design Overview}\label{design}

\subsection{Architecture}

\begin{figure}[htbp]
    \centering
    \includegraphics[width=0.75\linewidth, trim = 0.8cm 12.5cm 8.5cm 19cm, clip]{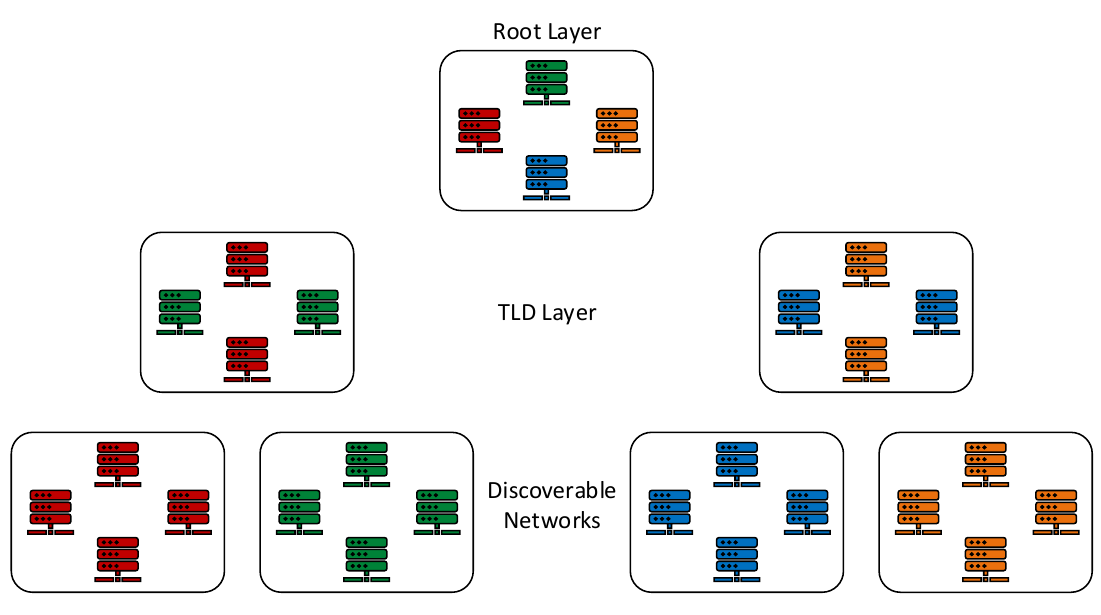}
    \caption{DNS Architecture Diagram}
    \label{fig:architecture}
\end{figure}

The main goals of this design are to: 
\begin{enumerate*}
    \item Maximize the decentralization of the discovery mechanism
    \item Ensure that the solution is scalable
    \item Allow for dynamic discovery services by enabling networks to voluntarily opt-in or opt-out.
\end{enumerate*}
To achieve this, we propose a multi-layered architecture for a blockchain-based discovery mechanism. Inspired by the traditional DNS architecture, blockchain network domains are organized into top-level domains (TLDs) as shown in Figure \ref{fig:architecture}. A root network manages the TLD networks by storing their connection information while remaining agnostic of individual domain details. 

Furthermore, the maintenance of the architecture is of great concern. This is because beneficiaries will use its services without contributing back unless otherwise incentivized. Therefore, the design ensures that networks interested in being discovered or having their assets discovered must actively contribute to maintaining the architecture. Specifically, a network seeking a DNS record must provide at least one node for its designated TLD network and another for the root network. The exact number of maintenance nodes required can be determined collectively by the participants in the network. Encouraged by the incentive mechanism discussed later, this design works towards maximizing the decentralization of the architecture, ensuring its scalability and providing redundancy to enable maximum up-time. 

\subsection{Architecture Operational Workflow}

\begin{figure}[htbp]
    \centering
    \includegraphics[width=\linewidth, trim = 12cm 17cm 14cm 17cm, clip]{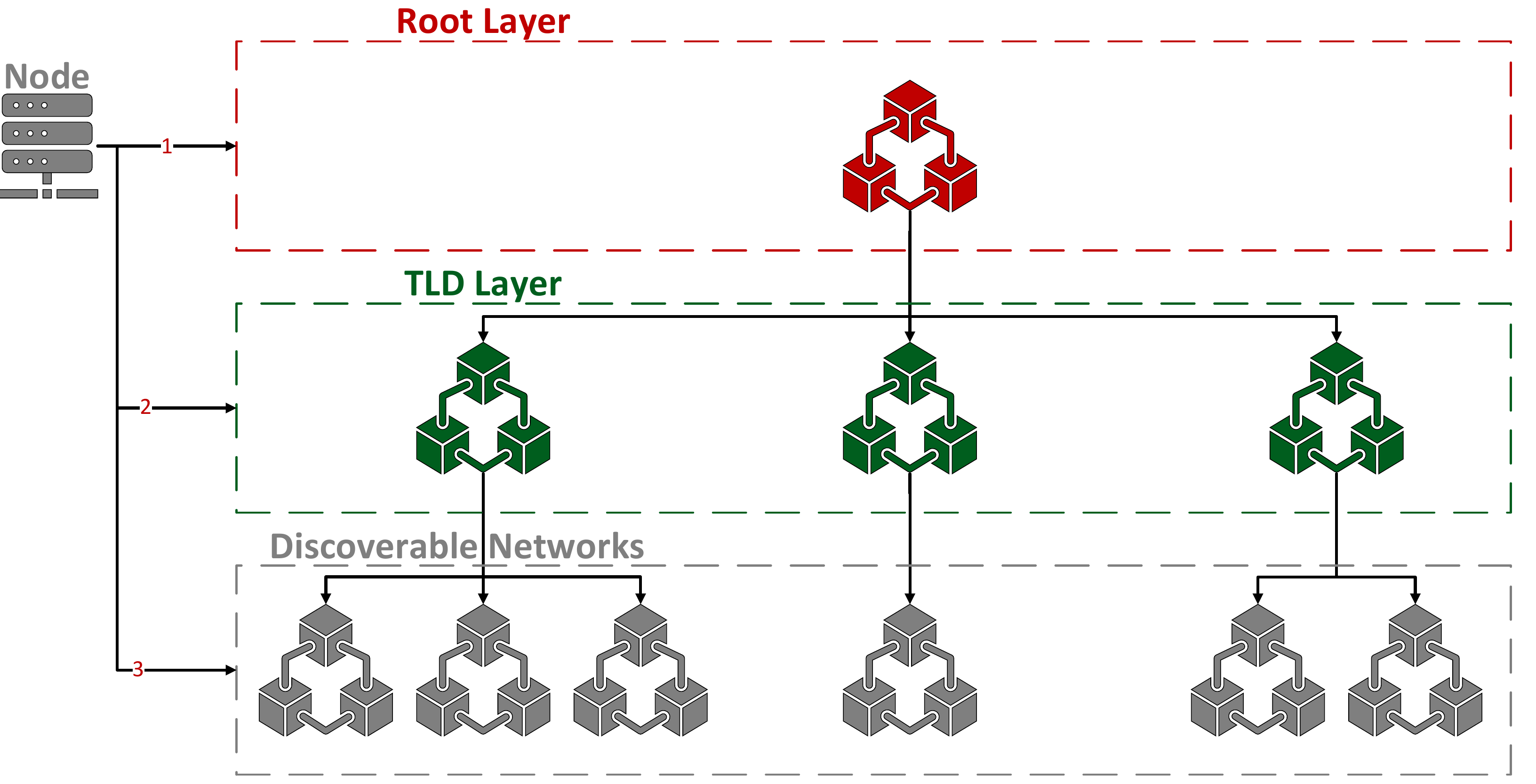}
    \caption{Domain Resolution Protocol}
    \label{fig:resolution_protocol}
\end{figure}

One of the main objectives of this design is to implement a dynamic discovery mechanism, eliminating the need for static lists of known networks and the reliance on centralized repositories. This dynamic approach allows DNS records to be updated automatically based on which networks choose to opt-in for discovery. The responsibility for registration is delegated to the networks themselves, bypassing the need for an administrator or governing entity to manually update the list of known networks. Instead, the architecture automatically processes registration requests and updates DNS records accordingly.

To understand how blockchain network domains can be registered within the architecture, it is essential first to explore how domains are resolved. Drawing inspiration from traditional DNS, domain name resolution follows a hierarchical process. In traditional DNS, domain names are resolved by a recursive resolver that queries root servers, TLD servers, and authoritative name servers in sequence to retrieve the IP address associated with a domain.

Similarly, as outlined in Figure \ref{fig:resolution_protocol}, a light client or an arbitrary node connects to the root network, whose connection details are static and well-known, and query the chain state for the connection details of the TLD network  associated with the domain. Using these details, it can then connect to the TLD network and query the connection information of the network associated with the domain it seeks to resolve. Finally, once resolved, the client establishes a direct connection with the target network, and these details can be cached locally for future use, improving efficiency.

For claiming domain names, traditional DNS architecture relies on domain registrars that offer reserved domains for a price determined by their popularity or rarity. However, in a decentralized environment, no single entity has control over a set of reserved domains. Instead, domains are claimed by blockchain networks on a first-come first-served basis, with networks only incurring the transaction fees associated with the registration process. Before registering a domain, a blockchain network must decide on the desired domain through its internal consensus mechanism. To claim a domain, a network must adhere to the following protocol:

\begin{enumerate}
    \item A node from the network, selected by its consensus mechanism, queries the root network to obtain the connection details of the TLD network associated with the desired domain.
    \item The selected node submits a transaction to the TLD network to register the domain. 
    \item Optionally, if the desired domain is already taken, the network can decide on an alternative domain and repeat the process.
\end{enumerate}


\subsection{Asset Discovery}

So far, the architecture has been discussed from the perspective of blockchain network discovery. While it is a crucial step toward realizing the vision of an Internet of Blockchains, asset discovery holds equal significance to facilitate efficient interaction across blockchains. 


The discovery architecture proposed in this study can be extended to include asset discovery functionality. The protocol outlines the process for a network to opt into asset discovery and to query the architecture to locate a specific asset. The primary goal of this design is to enable the inquirer to find a particular asset, provided they are already aware of its existence. This approach mitigates the risks associated with bad actors scraping the chain state to identify and exploit all registered assets on the network. By restricting discovery to known assets, the creation and registration of counterfeit or malicious assets is effectively discouraged. Furthermore, this design aligns with the precedent set by domain names, where an inquirer must be aware of a network's existence and know the associated domain name to interact with it.

\begin{figure}
\centering
\includegraphics[width=\linewidth, trim = 3cm 8cm 10cm 7cm, clip]{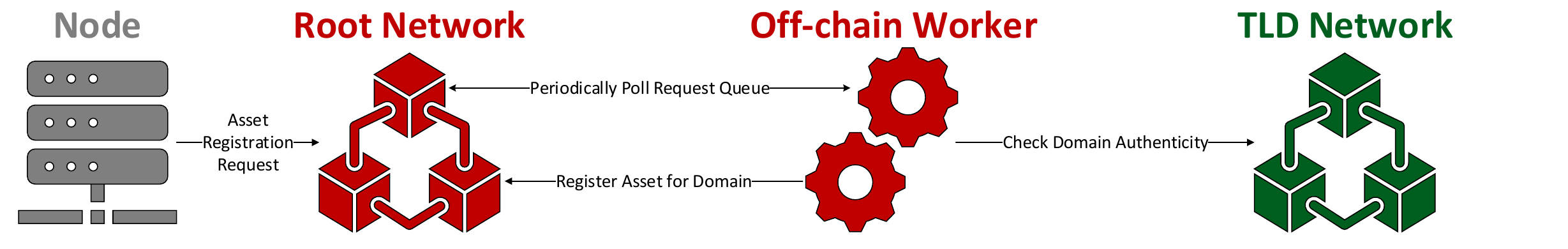}
\caption{Asset Registration Protocol}
\label{fig:asset_registration}
\end{figure}

To implement a similar strategy, the architecture stores hashes of asset identifiers rather than the raw identifiers themselves. The hashing process is delegated to the sender of the transaction -- using a previously agreed upon algorithm -- to ensure that the raw identifiers are not recorded in the transaction data. The assets are registered in the root network, as shown in Figure \ref{fig:asset_registration}. The registration process begins when a node submits a transaction to request the discovery registration of an asset. Once the transaction is received, the system creates a request and stores it in a queue on-chain for later processing. This deferred processing is necessary for two primary reasons:

\begin{enumerate}
    \item The incoming request must be validated to ensure that the domain is a legitimate domain registered within the architecture and that the requester is a maintainer (discussed in the incentive mechanism section) associated with the network requesting the asset registration.
    \item Transactions on the blockchain must be deterministic. However, the validation process requires external network requests, which are inherently non-deterministic.
\end{enumerate}

To address these requirements, off-chain workers are employed for the validation and processing of registration requests. These workers periodically poll the chain state to identify new requests added to the queue. Upon finding a new request, the worker retrieves and processes it. The system validates requests by first fetching the TLD connection details from the chain, and then utilizing the TLD network's RPC interface for the associated domain information. If the domain exists and the list of maintainers associated with it includes the identifier of the node that submitted the registration transaction, the worker executes a transaction to register the asset on-chain.

The asset registration structure uses a mapping approach where each asset serves as a key, and the corresponding value is a list of providers (domains) offering the asset. This design enables efficient querying, as an inquirer can retrieve a list of potential providers for a specific asset and proceed to interact with them directly. In addition, an inverse mapping is maintained on-chain, where each domain acts as a key, and its value is a list of assets it provides. This inverse mapping is crucial for enabling the removal of asset discoverability for specific providers. By knowing which assets a provider offers, the system can efficiently remove those assets from the provider's list when required. This dual-mapping structure ensures efficient querying and seamless management of asset discoverability. 

\section{Incentive Strategy}\label{incentive}

To achieve the goal of decentralization, the maintenance of this architecture relies on the networks that register themselves within it.  The root network and all TLD networks are designed to be composed of nodes belonging to the discoverable networks themselves. This approach ensures that no single entity has control over the discoverability of specific networks or the decisions made within the architecture, reinforcing its decentralized nature. However, from the perspective of registered networks, it is often more advantageous to simply use the architecture to make themselves discoverable without contributing resources to its maintenance. To address this, an incentive mechanism is introduced to ensure that all participating networks contribute to the system maintenance instead of simply free-loading off of others.

\begin{figure}
\centering
\includegraphics[width=\linewidth, trim = 20cm 16cm 5cm 7cm, clip]{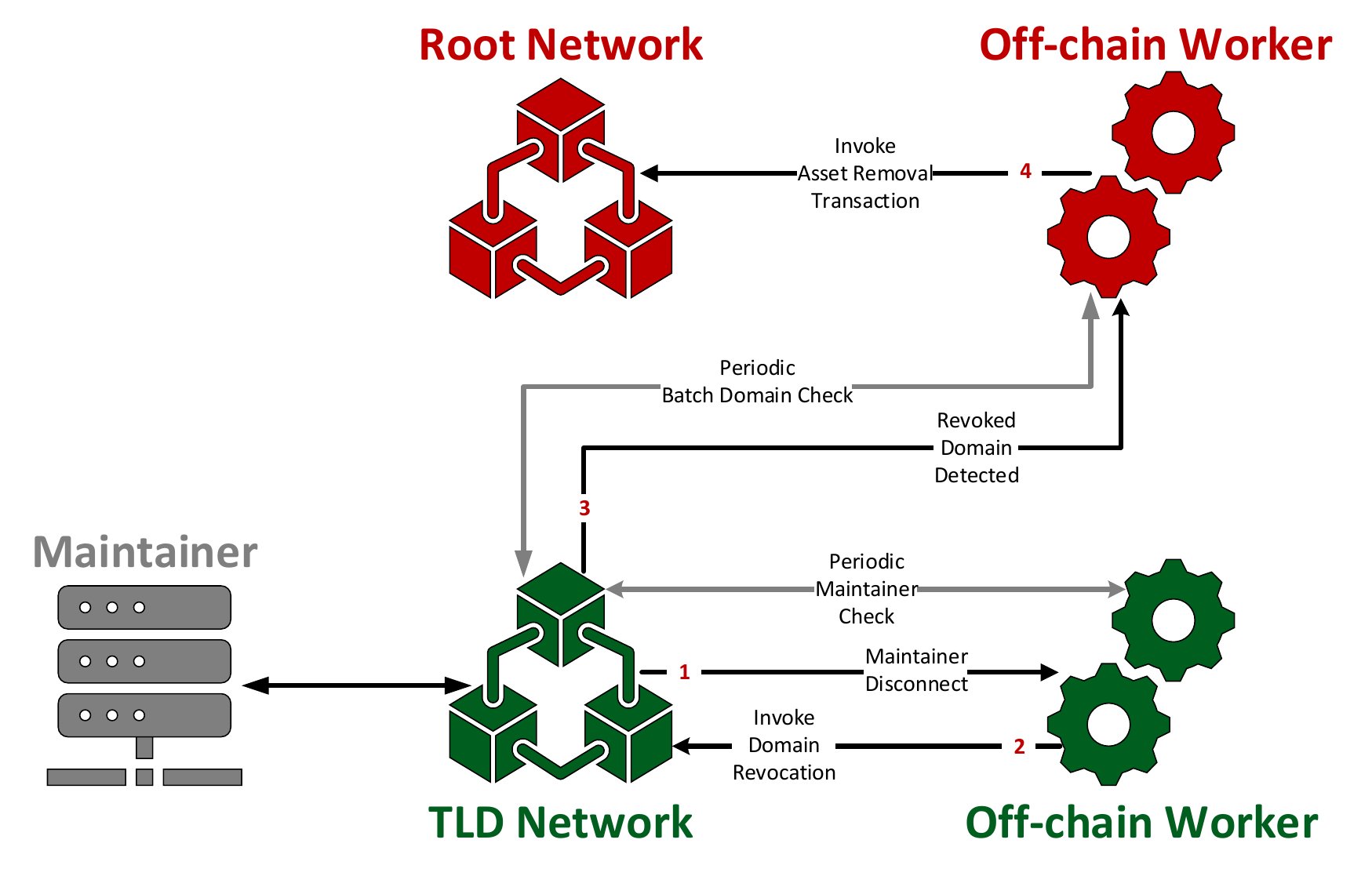}
\caption{Mandatory Participation Incentive Mechanism}
\label{fig:incentive_mechanism}
\end{figure}

In the proposed architecture, transaction fees serve as a primary incentive mechanism for the nodes maintaining the system. When a network is registered within the architecture, the registration is triggered through a transaction submitted to the architecture and is processed by the nodes maintaining it. Any operations related to the claimed domain such as revocation, amendment, or transfer, are also executed via transactions. These transactions generate fees, providing a direct financial incentive for nodes to maintain the architecture.

While transaction fees encourage the maintenance of the architecture, a more robust incentive mechanism is required to ensure the architecture remains operational in perpetuity. Figure \ref{fig:incentive_mechanism} shows the mandatory participation incentive mechanism utilized in this architecture. The goal of this mechanism is to establish a symbiotic relationship between the beneficiary networks and the discovery architecture. Specifically, any network that claims a domain within the architecture is required to contribute a minimum number of nodes to maintain both the TLD network to which it belongs and the root network.

This is achieved by having the networks provide a set of node identifiers which will be the maintainers provided by them to the architecture. Furthermore, the list of peers connected to the network is continuously monitored by having off-chain workers that poll the network periodically. If a node is detected to have left the network, the off-chain worker begins the process of revoking the domain associated with the node that ceased to maintain the network. This mechanism incentivizes networks to ensure the architecture never fails as long as they rely on the discovery services it provides.


\section{Implementation}\label{implementation}

The proposed architecture is implemented\footnote{The implementation of the architecture and its evaluation scripts can be accessed on: https://github.com/icdcs248/enabling-blockchain-interoperability-through-discovery} using the Substrate blockchain framework \cite{substrate2024}. Having been at the core of Polkadot \cite{wood2016polkadot}, Substrate is a well-tested empirically proven framework for creating and maintaining blockchain networks. Moreover, its implementation in the Rust programming language provides high performance for nodes, optimizing resource usage and execution speed. Substrate also supports Web-Assembly (WASM), enabling the development of lightweight clients that can run directly in web browsers. These clients facilitate seamless interaction with Substrate-based blockchains by querying chain states or submitting transactions for execution. This combination of performance, reliability, and accessibility makes Substrate an ideal choice for implementing the proposed architecture.

However, the main reason for selecting Substrate for this study's implementation lies in its modularity. It enables the modification of the blockchain runtime in an effortless manner through \textit{pallets}\footnote{https://docs.substrate.io/reference/frame-pallets/}. A Substrate network's runtime consists of numerous pallets integrated together in a plug-in/plug-out manner such that each pallet is responsible for a specific part of the functionality. For example, in Substrate's template node implementation\footnote{https://github.com/paritytech/substrate/blob/master/bin/node-template}, the block authoring mechanism is handled by \textit{pallet\_aura}, while the block finalization logic is managed by \textit{pallet\_grandpa}, together forming the network’s consensus mechanism. Accordingly, the architecture is implemented across three new distinct pallets, each designed to handle a specific functionality, as follows:

\subsection{The Root Pallet}
This pallet handles the core logic of the root network, which is primary responsible for managing the TLD networks and their connection information. Thus, this pallet maps the names of TLDs to the connection details required to interact with the network managing each TLD. In Substrate, this information is represented by \textit{chain specification} of a network, but the design is flexible to store connection details for any blockchain network built on any framework in a similar manner. Furthermore, Root pallet defines an extrinsic (transaction) named \textit{register\_tld} for the registration of a TLD network. When a registration request is submitted, the pallet checks whether the requested TLD is already managed by an existing network. If not, the pallet marks the requesting network as the TLD manager and maps them together in the on-chain storage.

\subsection{The TLD Pallet}
As the name suggests, this pallet handles the functionality provided by any TLD network within the architecture. Similar to the Root pallet, the TLD pallet maps the domains to their respective networks. However, for each domain, the pallet stores a \texttt{Domain Information} structure outlined in Table \ref{table:domain_info}, particularly containing a list of node identifiers referred to as \textit{maintainers}. These maintainers are the nodes provided by the network to support the maintenance of the TLD network in exchange for accessing the functionality of the discovery architecture.

\begin{table}[htbp]
    \centering
    \caption{Structure of \texttt{Domain Information}}
    \resizebox{\linewidth}{!}{%
    \begin{tabular}{| c | c |}
        \hline
        \textbf{Field Name} & \textbf{Description} \\
        \hline
        \texttt{Creator} & Contains the identifier of the account that claimed the domain. \\
        \hline
        \texttt{Chain Specifications} & The network connection information. \\
        \hline
        \texttt{Maintainers} & The list of identifiers of maintainer nodes supplied by the network. \\
        \hline
        \texttt{Available} & A boolean value indicating the domain's availability. \\
        \hline
    \end{tabular}%
    }
    \label{table:domain_info}
\end{table}

The pallet maintains an inverse map of maintainers mapped to the domain they are associated with. This is because it is more efficient to query the node identifier as the key of a map and get the associated domain in constant time rather than iterating over the list of domains along with the list of maintainers each domain provides, an approach whose time complexity depends on the number of domains registered in the TLD network. This inverse map is used to enforce the mandatory participation mechanism described in Section \ref{design}. 

\begin{algorithm}
    \caption{Mandatory Participation}\label{alg:mandatory_participation}
    \begin{algorithmic}
        \REQUIRE \textit{cached\_nodes}, \textit{current\_participants}, \textit{maintainer\_map}, \textit{revoke\_domain\_tx}
        
        \STATE $\textit{disconnected\_nodes} \gets \emptyset$
        
        \FOR{\textit{node\_id} in \textit{cached\_nodes}}
            \IF{\textit{node\_id} $\notin$ \textit{current\_participants}}
                \STATE $\textit{disconnected\_nodes} \gets \textit{disconnected\_nodes} \cup \textit{node\_id}$
            \ENDIF
        \ENDFOR
        
        \FOR{\textit{node\_id} in \textit{disconnected\_nodes}}
            \STATE $\textit{domain} \gets \textit{maintainer\_map}[\textit{node\_id}]$
            \IF{\textit{domain} $\neq \emptyset$}
                \STATE \textit{revoke\_domain\_tx(domain)}
            \ENDIF
        \ENDFOR
        
        \STATE $\textit{cached\_nodes} \gets \textit{current\_participants}$
    \end{algorithmic}
\end{algorithm}

The mandatory participation mechanism, shown in Algorithm \ref{alg:mandatory_participation}, is implemented in the pallet using off-chain workers. This approach is employed for two reasons:
\begin{enumerate}
    \item The network must be polled at regular intervals to verify the presence of nodes, ensuring compliance with the mandatory participation requirements.
    \item Blockchain transactions must be deterministic. Since polling peers for their presence involves network calls that are inherently non-deterministic in execution time, such operations cannot be included within a transaction.
\end{enumerate}
So, the off-chain worker executes the defined polling logic whenever a block is added to the chain. Polling is preferred over an event-based system due to its robustness against potential abuses. Nodes could exploit an event-driven system by dishonestly emitting events to feign compliance. Furthermore, even if a node is an honest participant, a sudden disconnection from the network -— caused by networking issues -— would fail to emit the necessary event to the architecture, leading to inaccuracies.

To circumvent this, the off-chain worker maintains a local cache on each node. This cache is used to store the identifiers of all nodes connected to the network. After a new block is authored, the off-chain worker compares the current list of nodes with the cached list from the previous block. If an identifier is missing, it indicates that the associated node has disconnected from the network. The off-chain worker then queries the on-chain map linking node identifiers to domains. If a match is found, it confirms that the disconnected node was a maintainer node associated with a domain, and the associated network's ownership of the domain is revoked.

The revocation process involves the off-chain worker calling the \textit{revoke\_domain} transaction, which is exclusively callable by the off-chain worker of the authoring node. Since domain information is stored on-chain, modifications must occur through transactions rather than off-chain methods. This transaction updates the domain record by setting its status to available, clearing its connection information, and allowing it to be claimed by another network. Additionally, the pallet provides two key transactions: \textit{register\_domain} for networks to claim available domains by mapping them to the requester's details, and \textit{amend\_chainspec} which enables domain owners to securely update their connection information after ownership verification.

\subsection{The Asset Discovery Pallet}
So far, the implementation has focused on providing the infrastructure necessary for network discovery. This pallet builds on the functionality provided by the other pallets to enable the discovery of assets across blockchain networks. To ensure direct and timely discovery of assets, this pallet is deployed on the root network of the architecture. Although this may initially appear to contradict the scalability-oriented design choices outlined in Section \ref{design}, the following justifications support this approach:
\begin{enumerate}
    \item Storing a mapping between domains and the assets they provide requires significantly less storage compared to storing connection information directly.
    \item The number of networks interested in exclusively making themselves discoverable is expected to be much larger than the number of networks making their assets discoverable.
\end{enumerate}
Accordingly, this pallet manages the information required for asset discovery directly on the chain of the root network.

Domains that provide discoverable assets are referred to as \textit{asset providers}. Asset identifiers are mapped to a list of potential providers on-chain, enabling responses to asset discovery queries with a list of possible providers. Networks interested in registering as asset providers must voluntarily submit requests specifying the assets they aim to make discoverable. Although these registrations are ideally managed via transactions, they require non-deterministic pre-processing, which cannot be executed within a transaction.

When a network submits a transaction to register its assets for discovery, two pieces of information must be verified:
\begin{enumerate}
    \item The transaction submitter must be a valid representative of the network, verified by matching their account identifier with the one used to register the domain.
    \item The provider’s domain must be a valid, active domain registered with the corresponding TLD network.
\end{enumerate}
This validation relies on information that is not directly accessible on the root network and must be retrieved from the provider's respective TLD network. To facilitate this, a specialized processing scheme is implemented.
\begin{table}[htbp]
    \centering
    \caption{Structure of \texttt{Pending Request}}
    \resizebox{\linewidth}{!}{%
    \begin{tabular}{| c | c |}
        \hline
        \textbf{Field Name} & \textbf{Description} \\
        \hline
        \texttt{Requester} & Contains the ID of the request initiator. \\
        \hline
        \texttt{Domain} & The domain name specified in the transaction initiating the request. \\
        \hline
        \texttt{Asset Hash} & The hash (using an arbitrary agreed upon algorithm) of the asset identifier. \\
        \hline
        \texttt{timestamp} & The block number at which the request was submitted. \\
        \hline
    \end{tabular}%
    }
    \label{table:pending_request}
\end{table}

The pallet exposes a \textit{register\_asset\_for\_domain} transaction, intended for use by external networks. This transaction creates a \texttt{Pending Request} on chain, format is shown in Table \ref{table:pending_request}. The pending request is then added to a request queue on the blockchain, and its timestamp is recorded to enforce an expiration period. This is required to regularly remove expired requests and prevent cluttering on-chain storage. Requests are deemed expired if the current block number exceeds a configurable \textit{lifetime} value. Once expired, they are removed from the blockchain's current state,  maintaining efficient storage utilization.

The processing of requests is handled by off-chain workers. Once a request is picked up by a worker,  it identifies the associated domain's TLD and queries the connection information for the TLD network from the chain. The worker then interacts with the TLD network through its Remote Procedure Call (RPC) interface. The storage key is first calculated by applying the Blake128 Concat hashing algorithm to the Simple Concatenated Aggregate Little-endian (SCALE) encoded domain string \cite{substrateKey}. Using the drived key, the worker constructs an RPC request to query the domain's information from the TLD network managing it. The TLD network responds with the \texttt{Domain Information} structure presented in Table \ref{table:domain_info}. The worker uses this data to verify the account identifier of the requester and confirm the availability of the domain. Upon successful verification, the worker triggers the \textit{submit\_verified\_domain} transaction, which updates the chain to list the domain as one of the asset's providers. 



        

Since domain name ownership can change, the availability of a domain is inherently volatile. To handle domain volatility, off-chain workers periodically verify provider status. An inverse provider-asset mapping enables efficient updates, while batched processing with persistent indexing ensures continuity between polling cycles, eliminating redundant checks. To enable persistent indexing, the index of the last accessed provider is maintained on-chain so that off-chain workers can continue where the previous cycle ended, rather than restarting the process.

\section{Discussions}\label{discussions}

\subsection{Decentralization}

Achieving decentralization can, in principle, be realized by implementing a DNS-like registry using any arbitrary distributed system equipped with a robust consensus mechanism. However, utilizing blockchain offers distinct advantages beyond basic decentralization. When configured with an appropriate consensus protocol and incentive mechanisms, blockchain inherently provides enhanced security and resilience through its decentralized architecture. This is primarily due to the immutability and transparency of the ledger, offered by the cryptographic techniques and consensus algorithms underpinning the blockchain. Furthermore, blockchain enables the integration of incentive mechanisms with direct monetary value, such as transaction fees, which encourage participation and sustain network integrity.

While implementing the DNS architecture as a single blockchain network satisfies the decentralization objective, it raises significant challenges for long-term viability. Registering all DNS records on a single chain results in growing storage demands as the number of records increases over time, increasing the maintenance demands and burdening the nodes.  Moreover, depending on the consensus mechanism employed, a single network may need to enforce limits on the number of validators (validator caps) to maintain efficiency. Restricting the number of validators will eventually constrain participation to a limited subset of nodes, leading to centralization. Therefore, this approach fails to achieve the second objective of the design: Scalability.

\subsection{Scalability}

The scalability challenges inherent in a single blockchain implementation are circumvented in this work by adopting a layered approach and distributing the architecture across multiple blockchain networks. In our design, each blockchain network manages a portion of the DNS information, thereby alleviating the storage constraints associated with maintaining all records on a single chain. This partitioning ensures that storage limitations are no longer a significant concern, enabling the system to scale as the number of records increases. The storage requirements are distributed as follows:

\begin{enumerate}
    \item The root network stores the connection information of the TLD networks. Since the number of TLD networks is significantly smaller than the total number of individual domains registered within the system, the root network's storage requirements will be minimal, and it will only be concerned with directing query requests to the appropriate TLD networks. 
    \item Each TLD network only stores information related to domains registered under its specific TLD. This design isolates the global domains into manageable subsets, reducing the storage demands on individual nodes within the network.
\end{enumerate}

Moreover, since each network operates with its own validator cap, nodes gain the flexibility to participate as validators across one or more blockchain networks responsible for domain resolution. By isolating domains into TLDs, each managed by an independent blockchain network, the architecture eliminates the need for a global validator cap. Instead, each network enforces its own validator limit, promoting a broader and more inclusive participation. Nodes have multiple opportunities to contribute to the discovery process by participating in the networks relevant to their domain: the root network and their respective TLD network. Additionally, the layered organization of the architecture into multiple TLD networks makes it easy to grasp, capitalizing on existing familiarity with traditional DNS.

\subsection{Dynamic Discovery}

The design enables dynamic discovery by allowing networks to voluntarily register themselves into the architecture. The necessary validations are then handled by the nodes in the architecture participating in its consensus. This eliminates the need for a central entity for accepting or rejecting potential registrations, resulting in a faster and more seamless process.

\section{Evaluation}\label{evaluation}

This section presents the evaluation results of the proposed architecture's implementation, focusing on its performance under heavy load conditions. As a network discovery service, the architecture is designed to handle a high volume of domain resolution queries efficiently. Timely response to these queries is critical to ensure the discovery process operates seamlessly, without introducing significant latency overhead that could hinder overall system functionality.

For the evaluation, a client application was created to facilitate the resolution of requested domains. Implemented in Golang, the client leverages the language's high performance and efficient thread management through Go routines. The client utilizes a library \cite{substrateGsrpc} to interact with the discovery architecture via Substrate's Remote Procedure Call (RPC) interface. The client implements the registration and resolution protocols detailed in Section \ref{design} for both network and asset discovery. Furthermore, it includes a benchmarking feature that allows the user to send a specified number of domain resolution requests per second. It spawns a new thread for each request and measures the time taken to receive a response for each request, enabling precise latency measurements.

To ensure an automated and systematic evaluation process, another Golang script was developed to manage the architecture setup. This script facilitates the creation of various network configurations with varying numbers of nodes to evaluate the scalability of the architecture. For each specified configuration, the script initializes the architecture by deploying the root network, the TLD networks, and the test networks, with all nodes running on Docker containers. The script then populates the architecture with data, adding the TLD information to the root network and registering the test networks' domain information with their respective TLD networks. Once the architecture is fully configured, the script uses the client's benchmarking feature to stress-test the system, evaluating its performance under heavy load conditions.

\subsection{Performance Evaluation}

\begin{figure*}
\centering
\includegraphics[width=\linewidth]{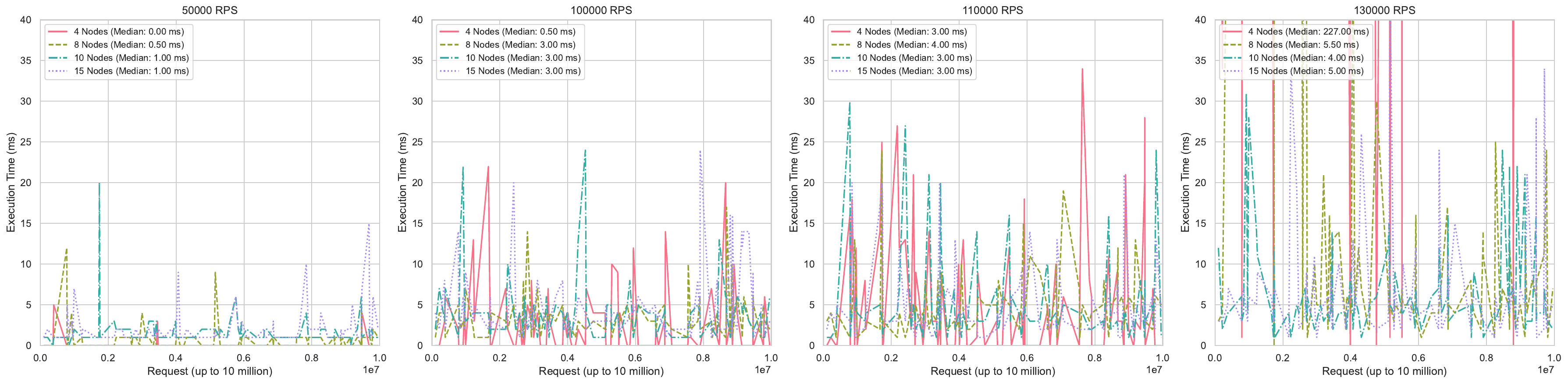}
\caption{Performance Evaluation Results}
\label{fig:eval}
\end{figure*}

The results of the evaluations are presented in Figure \ref{fig:eval}. All experiments were conducted on a machine equipped with an Intel Xeon Silver 4216 processor with 64 CPUs, 128 gigabytes of RAM, and 500 gigabytes of NVMe storage. The deployed architecture included the root network and two TLD networks, with each network configured to operate with 4, 8, 10, and 15 nodes in separate scenarios. For each configuration, a total of 10,000,000 requests were sent to the architecture, with concurrent request loads varying from 50,000 requests per second (rps) to 130,000 rps.

At the lower end of the load spectrum (50,000 rps), all configurations demonstrated extremely low latency, with median response times not exceeding 1 millisecond. As the request rate increased, the response times predictably rose across all configurations. For both the 100,000 and 110,000 rps tests, the median response times remained consistent between the two loads but were higher than those recorded at 50,000 rps. However, the 130,000 rps load shows significant performance degradation for the 4-node configuration, having a median response time of 227 milliseconds. This result is expected, as smaller networks become overwhelmed under high concurrent request volumes. However, the rest of the configurations were able to handle all 130,000 requests without issue with a maximum median latency of 5.5 milliseconds -- pointing to the scaling potential of the architecture that can be achieved by minimally increasing the number of nodes.

\subsection{Storage Evaluation}

The storage requirements for maintaining storing connection strings on the chain can be calculated as follows. Assuming the number of stored domains grow at a rate \textit{domain\_growth\_rate} per month, with an average size of \textit{string\_size} megabytes for each connection string. The storage requirement growth rate for a single-network architecture is given by:
\[storage\_growth\_rate = domain\_growth\_rate * string\_size \]
Assuming the number of TLDs deployed for the architecture is \textit{n}, and assuming an equal distribution of domains across these TLD networks, the storage growth rate per TLD network will be reduced by a factor of \textit{n} compared to the single-network solution. However, assuming an equal distribution of domains is not realistic. Concretely, the smallest connection string for a Substrate network in our tests was 767 kilobytes. With a growth rate similar to the current Internet infrastructure at approximately 244,000 domains daily \cite{circleid_domain}, a single network solution would exhibit a daily growth of about 182,761 megabytes. If domain distribution across TLDs follows a pattern similar to the current internet infrastructure, where the most utilized TLD accounts for 36.48\% \cite{domainstats}, the growth would only be 66,671 megabytes per day for the most heavily utilized TLD.

\section{Conclusion and the Future}\label{conclusion}

This paper introduced an inter-blockchain discovery architecture designed to enable dynamic and decentralized network and asset discovery to address a critical gap in achieving seamless interoperability for the Internet of Blockchains vision. Inspired by traditional DNS, the architecture integrates the root and TLD hierarchical model with blockchain networks, effectively separating concerns to create a decentralized, scalable, and robust discovery framework. To ensure the self-sustainability of the architecture, a robust incentive mechanism is designed and implemented. This mechanism mandates active participation from beneficiary networks by requiring them to contribute nodes to support the architecture's operation. This fosters a symbiotic relationship between beneficiary networks and the discovery framework, eliminating freeloading while maintaining a decentralized and resilient infrastructure. The architecture was implemented using Substrate blockchain framework and its performance was comprehensively evaluated. The evaluation results demonstrate the system's ability to handle a massive volume of concurrent domain resolution requests while scaling efficiently with the architecture's size. Furthermore, the architecture exhibited extremely low response times, validating its suitability for real-world deployment in high-demand scenarios.

The work presented in this study serves as a foundational step toward advancing the development of a cohesive Internet of Blockchains ecosystem while fostering discussion for further research in this direction. Several areas for improvement and exploration are identified: 

Trust and Certification: An important aspect that requires further exploration is trust. In this architecture, responses are inherently trusted because they are always returned from the chain, which is updated exclusively through consensus. However, enhancing response integrity by attaching a certificate is a valuable improvement. Investigating decentralized approaches to certificate generation presents an interesting opportunity to strengthen trust within the system.

 Node Accountability and Reputation: Leveraging disparate nodes to maintain the architecture presents an opportunity to strengthen accountability on a broader scale. Establishing a robust framework for measuring and reporting a global node reputation metric can provide networks with a valuable criterion for evaluating and onboarding new nodes. This approach promotes transparency and enhances reliability throughout the architecture.

Enhanced Asset Discovery: The current asset discovery mechanism employs a simple query structure. Future work could incorporate more complex asset characteristics, enabling more granular filtering and narrowing down of providers. This enhancement should aim to maintain the high query performance demonstrated in this study while improving the usability of the system for more intricate discovery scenarios.

Domain Squatting Mitigation: Domain squatting poses a potential challenge in the proposed architecture. Addressing this issue requires exploring mechanisms to verify the authenticity of networks claiming domains. For instance, implementing checks to ensure that test networks cannot be created solely for the purpose of claiming and squatting domains enhance the fairness and reliability of the discovery process.

\section*{Acknowledgment}

This work has been supported by a grant from the Web 3.0 Technologies Foundation which fosters all web3 technologies and in particular the Polkadot network.

\bibliographystyle{IEEEtran}
\bibliography{references}

\end{document}